\documentclass[aps,prb,epsfig,preprint]{revtex4}

\usepackage{graphicx}
\usepackage{color}
\usepackage{epsfig}
\usepackage{dcolumn}

\begin{document}

\title{Evolution of magnetism of Cr nanoclusters on a Au(111) surface}

\author{H. J. Gotsis and Nicholas Kioussis}
\affiliation{Department of Physics,
California State University Northridge,
Northridge, CA 91330-8268}
\author{D. A. Papaconstantopoulos}
\affiliation {Center for Computational Materials Science, Naval
Research Laboratory, Washington DC 20375}
 \affiliation {School of
Computational Sciences, George Mason University, Fairfax, VA 22030}

\date{\today}

\begin{abstract}

We have carried out collinear and non-collinear electronic
structure calculations to investigate the structural, electronic
and magnetic properties of isolated Cr atoms, dimers and compact
trimers. We find that the Cr monomer prefers to adsrob on the fcc
hollow site with a binding energy of  3.13 eV and a magnetic
moment of of 3.93 $\mu_B$. The calculated Kondo temperature of 0.7
K for the monomer is consistent with the lack of a Kondo peak in
scanning tunneling microscopy (STM) experiments at 7 K. The
compact Cr dimer orders antiferromagnetically and its bond length
contracts to 1.72 $\AA$ close to the value for the free-standing
Cr dimer. The very low magnetic moment of 0.005 $\mu_B$ for the Cr
atoms in the dimer is due to the strong $d-d$ hybridization
between the Cr adatoms. Thus, these calculations reveal that the
absence of the Kondo effect observed in STM experiments is due to
the small local moments rather than the Kondo quenching of the
local moments suggested experimentally. The Cr compact trimer
exhibits non-collinear co-planar magnetism with vanishing net
magnetic moment in agreement with experiment.
\end{abstract}

\maketitle

\section{Introduction}
\label{sec:introduction}

The study of magnetism at the nanoscopic scale is a major research
activity in condensed matter physics. Nanomagnetism is a highly
demanding fundamental problem as well as important for
applications in high density recording media and memory devices. A
particularly attractive feature of these systems is the strong
sensitivity of their electronic and magnetic properties to the
geometrical and chemical environment of the atoms. Consequently,
numerous experimental studies have been concerned with the
production and characterization of a large variety of nanometer
scale magnetic materials involving transition metals and noble
metals in different structural arrangements.\cite{NANO}

Magnetic transition metal nanostructures on non-magnetic substrates
have attracted recently large attention due to their novel and
unusual magnetic properties.\cite{cabria,himpsel} The supported
clusters experience both the reduction of the local coordination
number, as in free clusters, as well as the interactions with the
electronic degrees of freedom of the substrate, as in embedded
clusters, which may lead to the Kondo effect.\cite{hewson} The
complex magnetic behavior is usually associated with the competition
of several interactions, such as interatomic exchange and bonding
interactions, the Kondo effect, and in some cases noncollinear
effects, which can give rise to several metastable states close in
energy.\cite{himpsel,jing,oda} The ground state can therefore be
easily tuned by external action giving rise to the switching between
different states.

Recent advances in scanning tunneling microscopy (STM) and the
ability to build magnetic clusters with well-controlled interatomic
distances on metal surfaces have opened the possibility of probing
{\it local} interactions in magnetic nanostructures assembled atom
by atom at surfaces.\cite{crommie1,crommie2,crommie3} STM is ideal
for studying the low-energy structure near the Fermi energy (E$_F$)
since it allows access to states both above and below E$_F$. Recent
STM studies have investigated the interplay between magnetic and
electronic phenomena in one-impurity,\cite{crommie1}
two-impurity,\cite{crommie2} and more recently, in
three-impurity\cite{crommie3} systems. Cr is unique among the
transition-metal adatoms, because its half-filled valence
configuration (3$d^5$4$s^1$) yields both a large magnetic moment and
strong interatomic bonding leading to magnetic
frustration.\cite{cheng,kohl} The STM spectra for Cr clusters
supported on the Au (111) surface can be summarized as
follows.\cite{crommie3} The lack of a Kondo peak for the single Cr
atom was attributed to the fact that the Kondo temperature, T$_K$,
is significantly less than the experimental temperature of 7K. The
featureless STM spectrum observed for the dimer\cite{crommie2} was
suggested to be due to the strong antiferromagnetic coupling between
the atoms which in turn quench the Kondo effect. Interestingly,
compact triangular Cr trimers (Cr atoms occupying nearest-neighbor
sites on the (111) Au surface) exhibited two distinct classes of
behavior. In the first state, they displayed a featureless STM
spectrum, whereas in the second they displayed a narrow resonance at
the Fermi energy. Trimers were reversibly transferred from one state
to the other by very small shifts of one atom via tip manipulation.
It was suggested that the two observed states correspond to
equilateral and isosceles trimer configurations, with a net trimer
magnetic moment of zero and non-zero, respectively.

On the theoretical side, {\it ab initio} electronic structure
calculations have been employed to study the collinear magnetic
properties of 3$d$ adatoms and clusters on the (001)
Ag\cite{cabria,nonas} and Cu\cite{stepanyuk} surfaces and the (111)
Cu\cite{lazarovits} surface. The magnetic properties of small {\it
free} standing chromium clusters have been investigated by {\it ab
initio} electronic structure calculations for collinear\cite{cheng}
and noncollinear\cite{kohl} magnetic ordering and a wide range of
cluster geometries and atomic configurations. Both these
calculations give that the isosceles triangular structure is the
ground-state configuration, and that the noncollinear spins induce
only very small changes in the cluster's geometry relative to the
collinear model. Recently, the magnetic properties of noncollinear
structures of supported chromium clusters have been studied within
the phenomenological Anderson\cite{uzdin},
Coqblin-Schrieffer\cite{kudasov}and Heisenberg\cite{savkin} models.

The purpose of this work is to provide a first-principles
investigation of the structural, electronic, magnetic properties,
and the relative site preference of the Cr monomers, dimers, and
trimers adsorbed on the Au (111) surface. Even though the first
principles electronic structure calculations can not account for
the many-body resonance at E$_F$ associated with the Kondo
effect,\cite{kioussis} these calculations elucidate the origin in
the electronic structure of the evolution of the magnetic
properties (spin- and orbital- moments) with cluster size and
atomic geometry, including the possibility of noncollinear
magnetism. To the best of our knowledge, this is the first {\it ab
initio} study of noncollinear magnetic ordering for supported
equilateral triangular trimers.

The next section gives a brief description of the theoretical
method used in our study. In Sec. III we present the results of
the electronic structure and the magnetic properties of the
various Cr adatom configurations. Finally, we give a brief summary
in Sec. IV.

\section{Theoretical Method}
\label{sec:computation_methods}

The spin polarized electronic structure calculations were done
using the projector augmented-wave (PAW) method,\cite{PAW} as
implemented in the VASP code.\cite{VASP} For the
exchange-correlation potential we used the local spin-density
approximation (LSDA) functional of Perdew and
Wang.\cite{PERD-WANG} An energy cutoff of 260 eV was used for the
plane-wave expansion with the augmentation charge cutoff increased
to 1040 eV.

To model the Au(111) surface, we adopt the slab supercell
approach, where the slab consists of five gold layers with four
atoms per layer for the monomers and nine atoms per layer for the
dimers and the triangular trimers. A vacuum space corresponding to
four Au layers is used to separate the central slab and its
periodic images. The Brillouin zone integration was performed on a
Monkhorst-Pack 7x7x1 k-point mesh with a
Methfessel-Paxton\cite{METH-PAXT} smearing of 0.2 eV. The
calculations have been performed allowing relaxation of the top
three gold layers with the bottom two constrained at the bulk
geometry, with the calculated equilibrium lattice constant of
4.063 $\AA$ compared to the experimental value of $a$= 4.08 $\AA$.
Forces on the ions are calculated through the Hellmann-Feynman
theorem as the partial derivatives of the free energy with respect
to the atomic position, including the
Harris-Foulkes\cite{HARR-FOUL} like correction to forces. This
calculation of the forces allows a geometry optimization using the
quasi-Newton scheme.\cite{PULAY} Iterative relaxation of atomic
positions was stopped when the change in total energy between
successive steps was less than 0.001 eV. With this criterion,
forces on the atoms were generally less than 0.1 eV/$\AA$ with
lateral forces on chromium and gold surface layer atoms
negligible. The mass-velocity and Darwin correction terms are
incorporated into the PAW potentials and spin-orbit relativistic
effects are included self-consistently. While the spin-orbit
interaction determines the magnetocrystalline anisotropy energy,
it does not influence the size of the magnetic moments, presented
here, in a significant way. Non-collinear magnetic structures can
be treated at various levels of sophistication \cite{ncm} and the
PAW approach allows for a fully unconstrained vector-field
description of non-collinear magnetism.\cite{paw-ncm}

\section{Results and Discussion}
\label{sec:results_discussion}

\subsection{Monomers}

\begin{figure}
    \centerline{\resizebox{0.95\columnwidth}{!}{\includegraphics{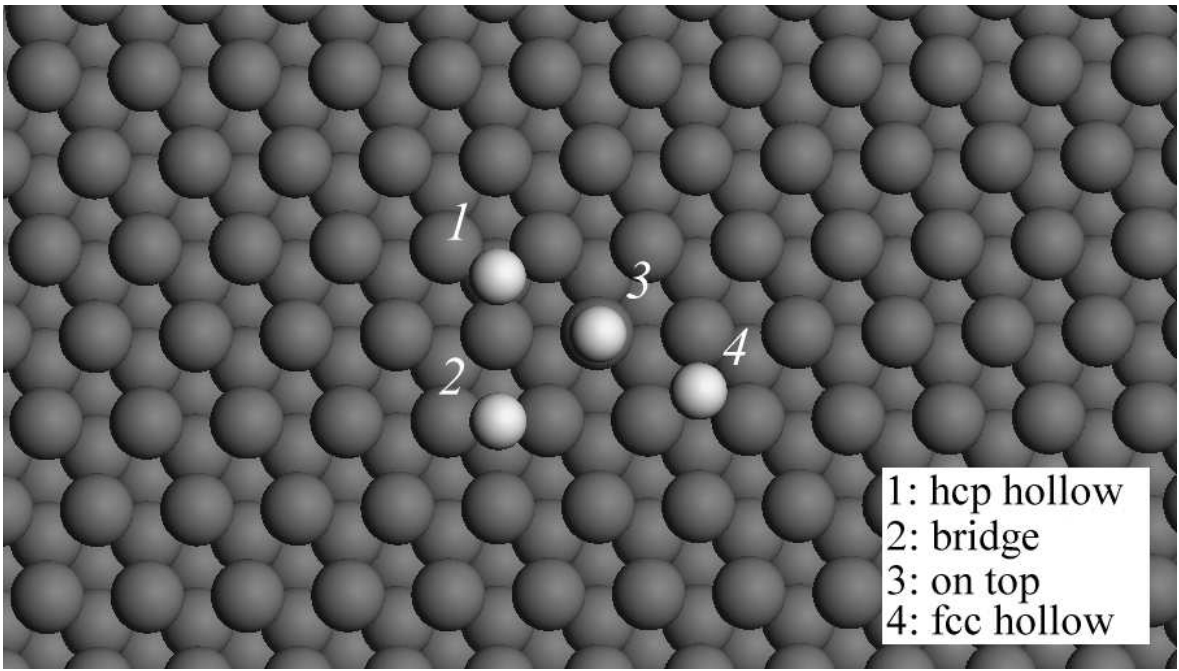}}}
    \vspace*{0.1in}

    \caption{Sketch of the Au(111) surface and its
    high-symmetry adsorption sites.}

    \label{fig:all_monomers}
\end{figure}

In order to understand the adsorption behavior of the Cr monomer,
we have calculated the binding energies, magnetic moments and
surface-atom distances for Cr adsorbed at four principal
adsorption sites of the Au(111) surface shown in
Fig.~\ref{fig:all_monomers}. These include on top sites, bridging
sites between two atoms, and hcp and fcc hollow sites between
three atoms.  In the hcp (fcc) hollow site there is an atom
directly beneath the second (third) layer. The adsorption energies
have been calculated using:

\begin{equation}
E_{bind} = E(Pure + Cr) - E(Pure) - E(Cr),
\label{equation:binding_energies}
\end{equation}
where $E(Pure + Cr)$ represents the total energy of the slab with
the Cr adatom, $E(Pure)$ represents the total energy of the pure
(clean) surface and $E(Cr)$ represents the total energy of an
isolated Cr atom. An isolated Cr atom was approximated by a single
atom in a simple cubic cell with lattice constant of 15 $\AA$. Its
total energy was then calculated using only the $\Gamma$-point in
the Brillouin zone integration. The binding energies, $E_{bind}$,
the spin magnetic moment, and the Cr-surface distance for the four
adsorption sites are listed in Table~\ref{table:monomers}. We find
that the preferred adsorption site is the fcc hollow, with a binding
energy of -3.129 eV and a spin moment of 3.93 $\mu_B$. Furthermore,
the binding energy on the hcp hollow site of -3.104 eV is very close
to that of the fcc hollow site, suggesting that the chromium binds
at high coordination sites. The on top site is less favorable by
about 0.8 eV and has the largest spin moment of 4.13 $\mu_B$. The
value of the magnetic moment for the on top adsorption site is in
good agreement with the calculated value of 4.3 $\mu_B$ for a Cr
adatom adsorbed on the top site on the Au (001) surface reported bu
Cabria {\it et al}.\cite{cabria} The larger value of the magnetic
moment in Ref. \cite{cabria} is due to the fact that lattice
relaxations were neglected in these calculations. We find that, as
expected, the spin moment is dominated by the contribution of the Cr
$d$-electron states (the $s$- and $p$- derived contributions are
less than $4\%$ of the total). Moreover, the induced spin moments on
the Au(111) surface are negligible. Note, that the trend of values
of the height of the Cr adatom from the Au surface, $d_{Cr-Au}$,
correlates well both with the trend of binding energies and the
magnetic moments, i.e., the shorter $d_{Cr-Au}$ is associated with
the largest binding energy and the smaller moment, due to the
stronger hybridization of the 3$d$-orbitals with the $6s$ conduction
band of the Au substrate.

We have also carried out spin-polarized electronic structure
calculations for the fcc hollow site with the spin-orbit coupling
included self-consistently. In Table~\ref{table:fcc_hollow} we
present the orbital- and spin- components of the magnetic moment
for a single Cr adatom, where the spin- and orbital- quantization
axis is chosen perpendicular to the surface. As expected, the
orbital-moment values for the adatom are low, consistent with the
zero orbital moment value for the isolated chromium atom. The
$S_z$ component of the moment is 3.903 $\mu_{B}$, which is
virtually identical to that from the calculation where the
spin-orbit coupling was not included (Table~\ref{table:monomers}),
indicating that the effect of the spin-orbit coupling is small.

\begin{figure}
    \centerline{\resizebox{0.95\columnwidth}{!}{\includegraphics{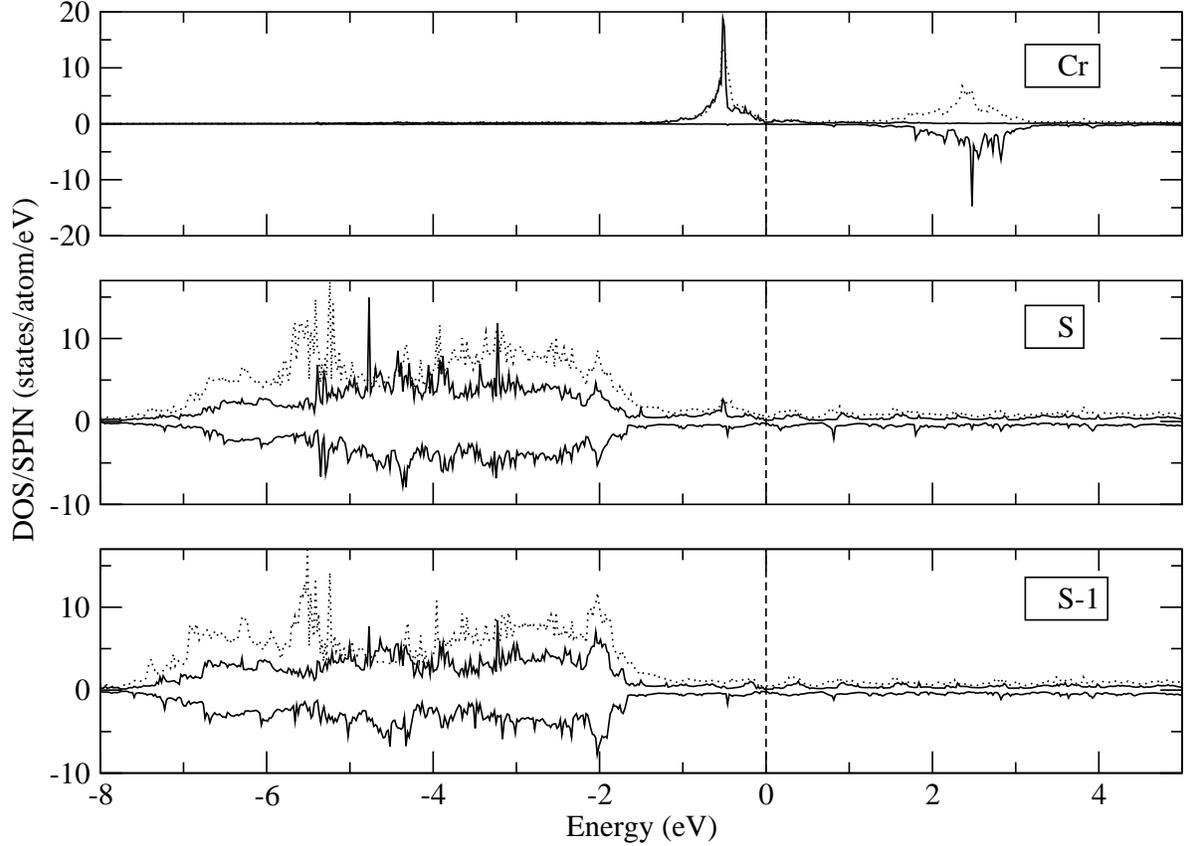}}}
    \vspace*{0.1in}

    \caption{Spin-projected local DOS of a Cr adatom on an fcc hollow
    site and the corresponding layer-resolved DOS of the
    Au(111) surface (S) and subsurface (S-1) (solid curves, respectively.
    The dotted curves denote the total DOS when including
    the spin-orbit coupling. The Fermi level is placed at 0 eV (dashed
    line).}

    \label{fig:monomer_DOS}
\end{figure}

In Fig.~\ref{fig:monomer_DOS} we show the spin-polarized local
density of states (DOS) of the Cr adatom on an fcc hollow site and
the layer-resolved DOS in the first two top layers (surface S and
subsurface S-1) of the five-layer Au(111) slab. The majority
(minority) spin states are shown in the positive (negative) portion
of each panel. The solid curves correspond to the spin-polarized DOS
without spin-orbit coupling and the dashed curves denote the total
DOS with spin-orbit coupling included. The Fermi energy, placed at 0
eV, lies in the $6s$ Au derived states region. The Cr adatom DOS for
the majority (minority) spin state shows Lorentzian-shaped virtual
states centered at $E_d - E_F$ ($E_d - E_F + U) \sim$ 0.5 eV ($\sim$
2.5 eV) below (above) the Fermi energy. This yields an intraatomic
Coulomb interaction $U$ for the Cr adatom of about 3 eV. The
majority spin virtual bound state has a half width of $\Delta \sim
0.075 eV$, due to the hybridization of the adatom $3d$ with the $6s$
states of the Au host. In the Anderson model which describes the
behavior of a paramagnetic impurity in a nonmagnetic metal host the
Kondo temperature $T_K$ can be calculated from \cite{hewson}
\begin{equation}
k_BT_K \sim \sqrt{\frac{\Delta U}{2}}
e^{-\frac{\pi}{2\Delta U}|E_d -E_F||E_d-E_F+U|}.
\end{equation}
Thus, we find that the single Cr adatom Kondo temperature is about $T_K \approx 0.5 K$,
which is much less than the experimental temperature of 7K, consistent with the featureless
STM spectrum observed experimentally.

\subsection{Dimers}

In order to study the evolution in electronic and magnetic
properties as two Cr monomers are merged into a single magnetic
molecule on Au(111), we have performed structural energy
minimization for a Cr dimer deposited on fcc hollow sites with an
unrelaxed dimer bond length of  2.87 $\AA$. As expected, the ground
state of the cluster was found to be antiferromagnetic, reminiscent
of the bulk. Interestingly, after relaxation of the chromium adatoms
and the outermost surface layers,  the dimer bond length contracts
to 1.72 $\AA$, which is consistent with the value of 1.68 $\AA$ for
the free Cr dimer reported by Cheng {\it et al}\cite{cheng} and
Bondybey {\it et al}\cite{BOND-ENGL}. At equilibrium, the two Cr
atoms occupy an hcp and an fcc hollow sites respectively,
 with moments of 0.005
and -0.005 $\mu_{B}$ for each adatom respectively. This is in
agreement with the value of 0$\mu_{B}$ per atom in the case of the
free dimer.\cite{cheng,kohl} For the dimer binding energy, a
generalization of Eq.~(\ref{equation:binding_energies}) gives the
value of -3.65 eV/adatom which is lower than the corresponding value
of -2.28 eV/atom for the free standing dimer.\cite{cheng} The low
magnetic moment and higher binding energy is due to the strong
$3d-3d$ hybridization between the Cr adatoms in the dimer which
results to a very stable electronic structure.\cite{cheng} Thus,
these calculations suggest that the absence of the Kondo effect for
the Cr dimer with bond length of 1.72 $\AA$ is due to the vanishing
of the magnetic moment of the adatom, rather than the experimentally
suggested\cite{crommie3} picture of non-zero Cr magnetic moments
which are locked in a singlet due to the antiferromagnetic exchange
interactions.

\begin{figure}
    \centerline{\resizebox{0.95\columnwidth}{!}{\includegraphics{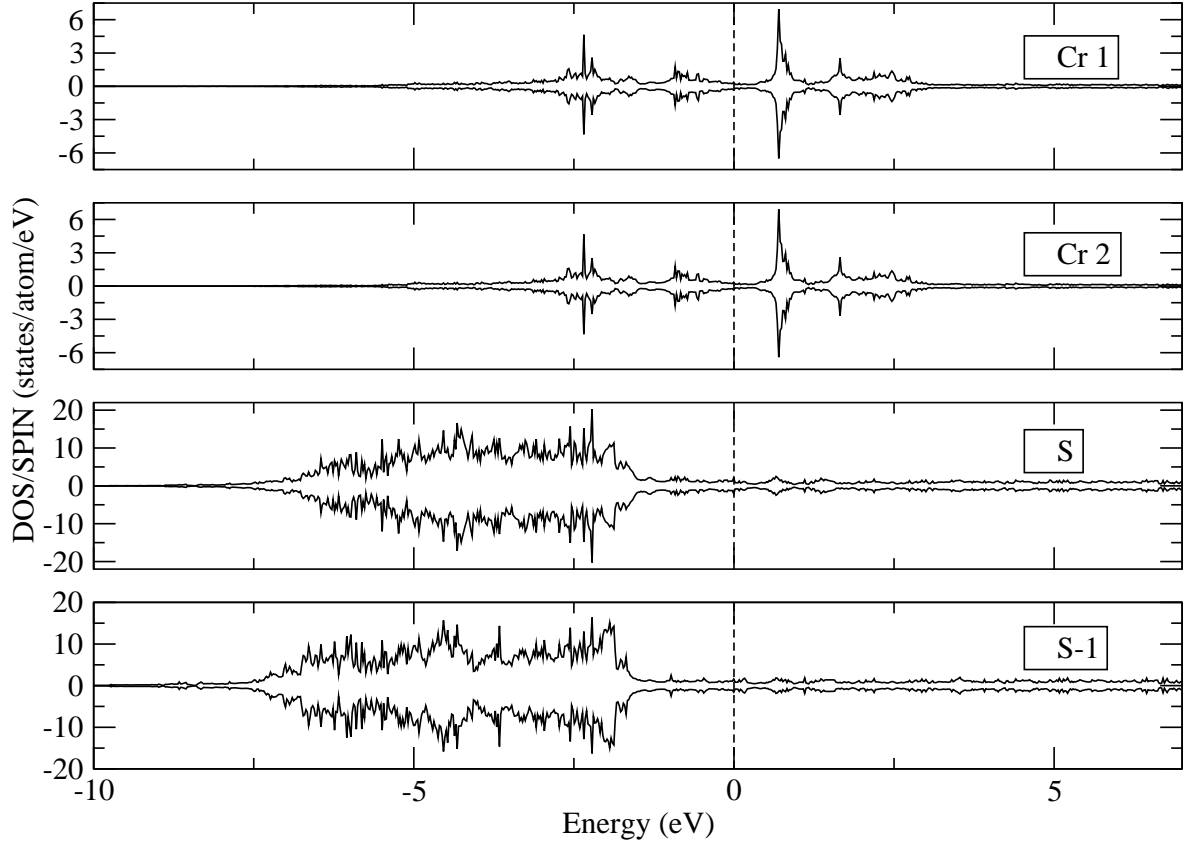}}}
    \vspace*{0.1in}

    \caption{Spin-projected local DOS of Cr as a dimer atom
    and the corresponding layer-resolved DOS of the
    Au(111) slab. The Fermi level is placed at 0 eV.}

    \label{fig:dimer_DOS}
\end{figure}

The spin-polarized (without spin-orbit coupling) local density of
states DOS  of the Cr adatoms in the dimer and the layer-resolved
DOS in the first two top layers of the Au(111) slab are shown in
Fig.~\ref{fig:dimer_DOS}. The two uppermost panels show the DOS of
the two Cr monomers. The majority (minority) spin states are shown
in the positive (negative) portion of each panel. For both Cr1 and
Cr2 adatoms, the two spin components are the same, and each adatom
is accordingly non-magnetic. This picture is consistent with the
values of the calculated magnetic moments. In comparison with
Fig.~\ref{fig:monomer_DOS} we find a broadening of the
Lorentzian-shaped virtual states and a shift of the $d$-orbital
states away from the Fermi energy ($E_d - E_F \sim$ 2.3 eV below
E$_F$) compared to the Cr monomer case. This results in a strong
hybridization between the chromium $d$ and gold $d$ bands. The
exchange interaction, $J_{sd}$, between the localized Cr-$d$
orbitals and the conduction electrons of the Au substrate can be
written in the form\cite{hewson},
\begin{equation}
J_{sd} \sim \frac{2V^2U}{(E_d-E_F)(E_d+ U_d-E_F)},
\end{equation}
where, $V$ is the $d-$conduction electrons hybridization. Thus, the
shift of the $d$ orbital states
 away from $E_F$ for the Cr atom in the dimer leads to
a reduction in $J_{sd}$ and hence, a reduction of the Kondo
temperature, $T_K$, for Cr dimers.

\subsection{Trimers}

Our {\it ab initio} calculations for collinear antiferromagnetic
ordering perpendicular to the surface failed to converge due to the
magnetic frustration associated with the geometry of the equilateral
trimer. On the other hand, we find that the compact Cr trimer in the
equilateral configuration displays coplanar non-collinear spin
antiferromagnetic  ordering due to the competition of the exchange
interactions between the different atoms. The electronic and
magnetic structures are investigated within non-collinear magnetism
by considering each magnetic moment on each Cr atom as a vector.  We
have performed structural energy minimization for a Cr equilateral
trimer originally deposited on fcc hollow sites with a bond length
of 2.87 \AA (nearest-neighbor sites). In the equilibrium
configuration, the bond length contracts to 2.39 $\AA$ preserving
the equilateral geometry. From Table~\ref{table:trimer} we observe
that the angle between each pair of moments equals $120^{\circ}$ and
the total spin moment of the trimer is zero.
 The modulus of the
moments on each atom of the trimer of 3.15 $\mu_B$ is comparable to
the monomer values and remarkably enhanced relatively to the bulk
spin moment value of approximately 0.5 $\mu_{B}$/atom\cite{fawcett}
(Cr bulk has an antiferromagnetic ground state with a moment smaller
than might be expected from its half filled 3$d$ shell, indicating
its nearness to a magnetic instability).

\begin{figure}
    \centerline{\resizebox{0.95\columnwidth}{!}{\includegraphics{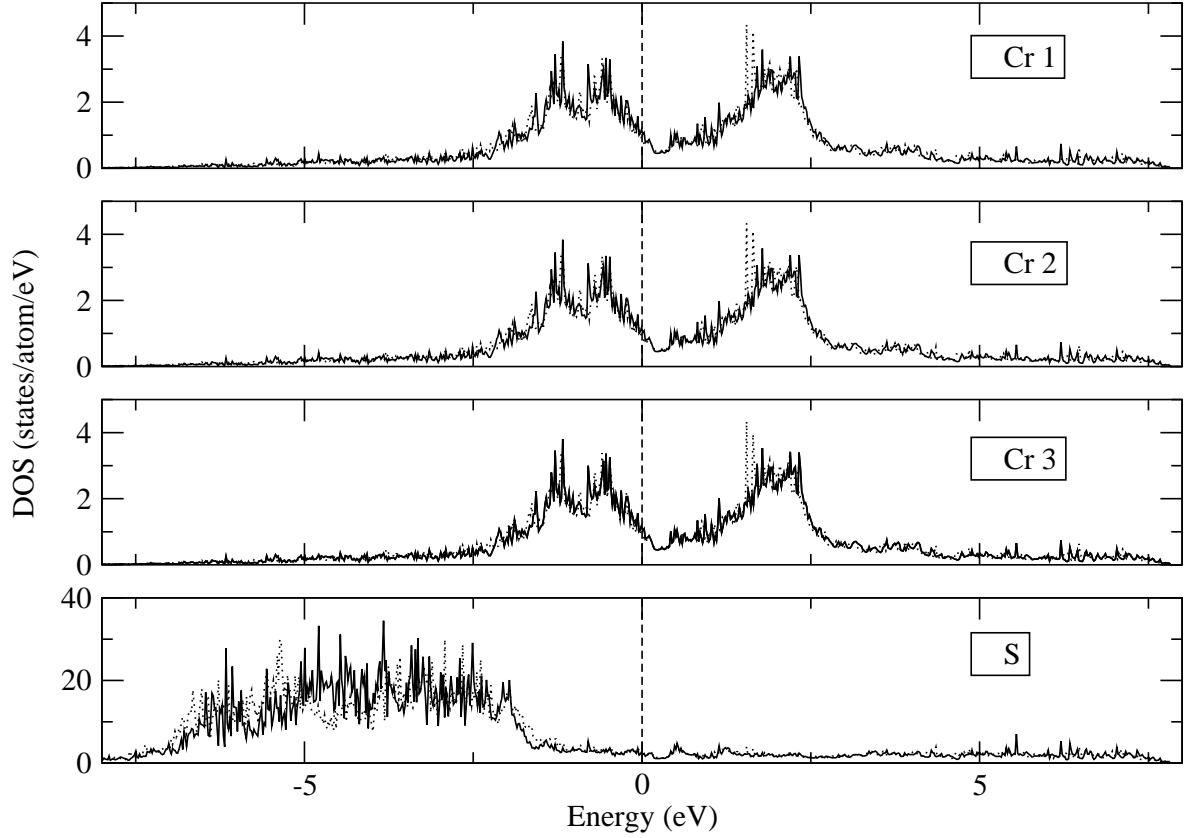}}}
    \vspace*{0.1in}

    \caption{Local DOS of Cr as a trimer atom in a non-collinear
    configuration and the corresponding surface layer DOS of the Au(111) slab without
    spin-orbit coupling (solid curve). The dotted line denotes the total DOS when spin-orbit
    coupling is included. The Fermi level is set at 0 eV.}

    \label{fig:trimer_DOS}
\end{figure}

Figure~\ref{fig:trimer_DOS} shows the local DOS of the Cr adatoms
and at the Au atoms in the surface. The DOS of Cr atoms are all
the same due to the symmetric structure. With the atomic moments
deviating from collinear alignment, the local spin projection is
no longer a good quantum number specifying the electron states as
spin-up and spin-down. However, due to the large magnetic moment,
in the energy region below (above) the Fermi level there are
preferentially states with a positive (negative) spin projection.
The strong hybridization between the Cr $d$ orbitals leads to a
wider Cr bandwidth and the splitting of the Lorentzian-shaped
virtual state for the monomer in Fig.~\ref{fig:monomer_DOS}.

It is known that, besides exchange, the spin-orbit coupling (SOC)
can also be responsible for noncollinearity of the magnetic
structure even if the SOC is very small in the 3d transition
elements. Therefore, we carried out calculations that take account
of both the noncollinearity of the magnetic moments and the SOC. One
can see in Fig. ~\ref{fig:trimer_DOS} that the effect of SOC on the
DOS of Cr atoms is small. In Table ~\ref{table:trimer_soc} we list
values of the orbital and spin components of the magnetic moments
for the Cr trimer. The orbital moments are in general a factor of
two larger than the those for the monomer orbital
(Table~\ref{table:fcc_hollow}), whereas the total magnetic moments
$M$ are smaller than the monomer case. Finally, the spin magnetic
moments are very close to the values from the calculation where the
SOC was not included, indicating that the magnetic (spin only)
frustration is the main reason for the non-collinear structure
formation in the trimer (Table~\ref{table:trimer}).

\section{Conclusion}

In conclusion, we have carried out first-principles electronic
structure calculations to study the structural, electronic and
magnetism properties of monomer, dimer, and trimer Cr clusters on
Au(111) surface. Our results show that the fcc hollow is the most
favorable location for the single Cr adatom and has a binding energy
of 3.3 eV and a magnetic moment of 3.93 $\mu_B$. The Cr DOS for the
majority (minority) spin state shows Lorentzian-shaped virtual
states centered at $E_d - E_F$ ($E_d - E_F + U) \sim$ 0.5 eV ($\sim$
2.5 eV) below (above) the Fermi energy. The calculated single Cr
Kondo temperature of 0.7 K is consistent with the lack of a Kondo
effect at 7 K in STM experiments. The compact dimer is found to
order antiferromagnetically, has a very short bond length of 1.72
$\AA$ and a very low magnetic moment of 0.005 $\mu_B$. This is due
to the strong $d-d$ hybridization between the Cr atoms. Thus, these
calculations reveal that the lack of the Kondo effect observed in
STM experiments is due to the small Cr magnetic moments rather than
the Kondo suppression of the local moment, as has been suggested in
the analysis of experiments. The results for the non-collinear
triangular compact trimer predict that the net spin cluster magnetic
moment is zero, consistent with the featureless spectrum of the STM
study. Future work will be aimed at investigations of the linear
trimer and isosceles triangle cases, as well as of collinear
magnetism perpendicular to the surface.

\section*{Acknowledgements}
One of us (H.J.G.) would like to thank Gang Lu and N. C. Bacalis for
useful conversations and advice. This work was supported in part
under NSF under Grant No DMR-00116566 and US Army under Grant No
AMSRD-45815-PH-H and W911NF-04-1-0015, respectively.

\begin{table}
    \caption{Calculated binding energies $E_{bind}$, spin magnetic moments and
    Cr-surface distances $d_{Cr-Au}$ for Cr adatoms adsorbed on different sites
    on the Au(111) surface.}
\begin{ruledtabular}
    \begin{tabular}{llll}
        Adsorption site  & $E_{bind}$ (eV)  & Magnetization ($\mu_{B}$)  & $d_{Cr-Au}$ ($\AA$)\\ \hline
    on top       & -2.334           & 4.128                      & 2.426 \\
        fcc hollow       & -3.129           & 3.929                      & 1.931 \\
        bridge           & -3.046           & 3.933                      & 1.993 \\
        hcp hollow       & -3.104           & 3.939                      & 1.937
    \end{tabular}
\end{ruledtabular}

    \label{table:monomers}
\end{table}

\begin{table}
    \caption{Calculated magnetic moments
    for a Cr adatom in an fcc hollow at the Au(111) surface. The orbital magnetic
    moments $L_{\delta}$ for $\delta$= $x$,$y$,$z$, the corresponding
    spin magnetic moments $S_{\delta}$ and the total magnetic moment $M$ are
    given in Bohr magnetons ($\mu_{B}$).}
\begin{ruledtabular}
    \begin{tabular}{lllllll}
        $L_{x}$  & $L_{y}$  & $L_{z}$  & $S_{x}$  & $S_{y}$  & $S_{z}$  & $M$\\ \hline
    -0.011   & -0.011   & -0.010   & 0.018    & 0.003    & 3.903    & 3.893
    \end{tabular}
\end{ruledtabular}

    \label{table:fcc_hollow}
\end{table}

\begin{table}
    \caption{Local spin moments of a equilateral Cr trimer, in a
    non-collinear magnetic configuration, deposited on Au(111)
    surface. The magnetic moments and the modulus of the moments
    are given in Bohr magnetons ($\mu_{B}$).}
\begin{ruledtabular}
    \begin{tabular}{lllll}
        Cr site   & $S_{x}$  & $S_{y}$  & $S_{z}$  & Modulus\\ \hline
    Cr1   & -0.005   &  3.157   & -0.030   & 3.157 \\
        Cr2       & -2.736   & -1.578   & -0.008   & 3.158 \\
        Cr3       &  2.728   & -1.578   & -0.003   & 3.152
    \end{tabular}
\end{ruledtabular}

    \label{table:trimer}
\end{table}

\begin{table}
    \caption{Calculated magnetic moments of a equilateral Cr trimer, in
    a non-collinear magnetic configuration, at Au(111) surface. The orbital magnetic
    moments $L_{\delta}$ for $\delta$= $x$,$y$,$z$, the corresponding
    spin magnetic moments $S_{\delta}$ and the total magnetic moment $M$ are
    given in Bohr magnetons ($\mu_{B}$).}
\begin{ruledtabular}
    \begin{tabular}{llllllll}
      Cr site   & $L_{x}$  & $L_{y}$  & $L_{z}$  & $S_{x}$  & $S_{y}$  & $S_{z}$  & $M$\\ \hline
      Cr1   &  0.000   & -0.027   &  0.024   & -0.005   &  3.175   & -0.027   & 3.148 \\
      Cr2       &  0.024   &  0.014   &  0.024   & -2.751   & -1.587   & -0.005   & 3.148 \\
      Cr3       & -0.024   &  0.014   &  0.024   &  2.744   & -1.587   &  0.000   & 3.142
    \end{tabular}
\end{ruledtabular}

    \label{table:trimer_soc}
\end{table}

\end{document}